\documentclass[flushrt,preprint]{aastex}
\usepackage{amssymb}
\usepackage{amsmath}
\usepackage{graphicx}
\usepackage{psfrag}
\begin{document}
\psfrag{aaD}{$a D$}
\psfrag{aaoD}{$(1-a)D$}
\psfrag{ph}{$\phi$}
\psfrag{dt}{$\Delta\theta$}
\psfrag{pt}{$\pi-\Delta\theta$}
\psfrag{ch}{$\chi$}
\title{Implications of Fast Radio Burst Pulse Widths}
\shorttitle{FRB Pulse Widths}
\shortauthors{Katz}
\author{J. I. Katz\altaffilmark{}}
\affil{Department of Physics and McDonnell Center for the Space Sciences}
\affil{Washington University, St. Louis, Mo. 63130}
\email{katz@wuphys.wustl.edu}
\begin{abstract}
The pulse widths, dispersion measures and dispersion indices of Fast Radio
Bursts (FRB) impose coupled constraints that all models must satisfy.  We
show that if the dispersion measures resulted from propagation through the
intergalactic medium from cosmological distances and the pulse widths were a
consequence of scattering by single thin screens, then the screens' electron
densities were $\gtrsim 20$/cm$^3$, $10^8$ times the mean intergalactic
density.  This problem is resolved if the radiation scattered close to its
source, where high densities are plausible.  Observation of dispersion
indices close to their low density limit of $-2$ sets a model-independent
upper bound on the electron density and a lower bound on the size of the
dispersive plasma cloud, excluding terrestrial or Solar System origin.  The
scattering and much of the dispersion may be attributed to regions about 1
AU from the sources, with electron densities $\sim 3 \times 10^8$ cm$^{-3}$.
The inferred parameters are only marginally consistent; re-examination of
the assumed relation between dispersion measure and distance is warranted.
Origin in an ionized starburst or protogalaxy is suggested, but statistical
arguments exclude compact young SNR in the Galactic neighborhood.  An
appendix applies these arguments to PSR J1745-2900 at the Galactic Center.
We suggest that its pulse width and angular broadening may be reconciled if
we are near a caustic or focal point produced by refraction, rather than by
the classic thin sheet scattering model.
\end{abstract}
\keywords{radio continuum: general --- intergalactic medium --- plasmas ---
scattering}
\maketitle
\section{Introduction}
\cite{T13} discovered four fast radio bursts (FRB) whose large dispersion
measures (DM) and high Galactic latitudes indicated that their sources were
at cosmological distances.   FRB 110220 had an observed dedispersed width
$W = 5.6 \pm 0.1$ ms (at a frequency $\nu = 1300$ MHz), while only upper
limits on $W$ were found for the remaining three FRB.  \cite{BSB14}
discovered FRB 011025 for which $W = 9.4 \pm 0.2$ ms.  Fitting $W \propto
\nu^\beta$,  both these FRB had scattering indices $\beta$ in agreement with
the predicted $\beta = -4$ for multipath propagation spreading in a
refractively scattering plasma medium.  Two other FRB, 010621 \citep{K12}
and 121102 \citep{S14b}, had measured widths but these widths were not
attributed to scattering; these FRB occurred at low Galactic latitudes,
hinting that they may be Galactic.  We do not discuss them explicitly, but
their parameters are similar to those of FRB 110220 and FRB 011025, with
similar implications if the same assumptions are made.

This paper explores the implications of the assumptions that the dedispersed
pulse widths of FRB are a consequence of scattering in intergalactic plasma
and that the dispersion measures indicate cosmological distances.  Any
explanation of these observations must account for two facts: (1) All FRB
have dispersion measures within a range of a factor of about two (three if
the Lorimer burst \citep{L07} is accepted as an FRB), implying a universal
property, not an unusual circumstance such as a line of sight that happens
to intersect a rare dense cloud; (2) The dispersion index is very close to
$-2$, and consistent with exactly $-2$, implying an upper bound on the
density of the dispersing plasma and a lower bound on its size.  These facts
are readily accounted for if dispersion occurs in the intergalactic medium,
but this appears inconsistent with the pulse broadening, interpreted as the
result of scattering.

Section~\ref{wherenot} presents our central result, that the scattering
responsible for the pulse widths of these FRB did not occur in the general
intergalactic medium.  Section~\ref{where} discusses where the scattering
may have occurred.  Section~\ref{dindex} obtains limits on the plasma
density in the scattering region that can be inferred from the observed
dispersion indices.  Section~\ref{parameters} sets bounds on the parameters
of the scattering region.  Section~\ref{others} applies these arguments to
Galactic PSR, including the heavily broadened and dispersed PSR J1945--2900
at the Galactic Center.
Section~\ref{implications} considers the implications for models of FRB.
Section~\ref{discuss} contains a concluding discussion.  Because pulse
widths have been measured for FRB1100220 and FRB011025, we present numerical
results as ordered pairs (110220, 011025).
\section{Pulse Widths}
\label{wherenot}
\begin{figure}[!h]
\centering
\includegraphics[width=1.5in]{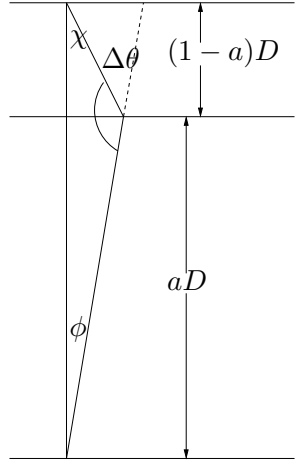}
\caption{\label{pathfig}Path of scattered radiation}
\label{propagation}
\end{figure}

We make the approximation that FRB (110220, 011025) were at distances $D =
(2.8,2.2)$ Gpc \citep{T13,BSB14} in a flat static universe.  For the
estimated redshifts $z = (0.81,0.61)$ this only introduces an error of a
factor ${\cal O}(1)$, less than other uncertainties.  Following the classic
theory of \cite{W72}, we approximate the
propagation paths as produced by a single scattering at a distance $aD$ from
us and $(1 - a)D$ from the source.  If the scattering angle $\Delta \theta
\ll 1$ then the angles $\phi \approx (1 - a) \Delta \theta$ and $\chi
\approx a \Delta \theta$; the geometry is shown in Fig.~\ref{propagation}.

We assume that the origin of the pulse width $W$ is dispersion in
propagation path lengths.  The total propagation delay corresponding to the
reported (after subtracting estimated Galactic contributions) $\mathrm{DM}
\approx (910,680)$ pc-cm$^{-3}$ \citep{T13,BSB14}
\begin{equation}
\Delta t_{DM} = {2 \pi e^2 \over m_e c \omega^2} \mathrm{DM} \approx 
(2.2,1.7)\,\mathrm{s} \approx (400,180) W
\end{equation}
at $\nu = 1300$ MHz, where the last approximate equality compares the
$\Delta t_{DM}$ calculated from the empirical DM to the empirical $W$.  This
assumption implies an assumption about the homogeneity of the intervening
medium, in which most of the dispersion is presumed to originate, on scales
${\cal O}(\Delta \theta D) \ll D$ of the separation, perpendicular to the
propagation direction, of the weakly scattered paths.  

The incremental delay attributable to scattering by an angle $\Delta \theta$
(\cite{W72}; \cite{Ku14}) is
\begin{equation}
\label{delay}
W \approx {D \over 2 c} (\Delta \theta)^2 a (1 - a).
\end{equation}
Then
\begin{equation}
\label{Dtheta}
\Delta \theta \gtrapprox \sqrt{8 c W \over D} \approx (4 \times 10^{-10},
6 \times 10^{-10}),
\end{equation}
the minimum value obtained for $a = 1/2$.  The angular width of the
received radiation
\begin{equation}
\label{phi}
\phi \approx (1 - a) \Delta\theta = \sqrt{{2 c W \over D}
\left({1-a \over a}\right)}.
\end{equation}

Refraction by a surface whose normal is tilted from the direction of
propagation by an angle $\theta$, with a ratio $n$ of refractive indices
between the two sides of the surface, leads to a deflection, unless $\vert
\pi/2 - \theta \vert \lesssim {\cal O}(\sqrt{\vert 1 - n \vert}) \ll 1$,
\begin{equation}
\label{Dthetan}
\Delta \theta \approx \vert 1 - n \vert \tan\theta.
\end{equation}
In general $\tan\theta = {\cal O}(1)$.  Taking this as an approximate
equality and writing $2 \pi \nu \equiv \omega \gg \omega_p$
\begin{equation}
\label{neq}
4 \times 10^{-10} \lessapprox \Delta \theta \approx 1 - n \approx
{1 \over 2} {\delta \omega_p^2 \over \omega^2}.
\end{equation}
From the expression for the plasma frequency
\begin{equation}
\label{omegap}
\delta \omega_p^2 = {4 \pi \delta n_e e^2 \over m_e},
\end{equation}
where $\delta n_e$ is the magnitude of fluctuations in the electron density,
we find 
\begin{equation}
\label{ne}
\delta n_e \approx {m_e \omega^2 \over 2 \pi e^2} \sqrt{2 c W \over a (1-a)
D} \gtrapprox n_{e,min} \equiv {m_e \omega^2 \over 2 \pi e^2} \sqrt{8 c W
\over D} \approx (17,24)\,\mathrm{cm}^{-3}.
\end{equation}
An elementary calculation shows that (\ref{ne}) also applies to the
refractive (dispersion in group velocity) delay if the path is heterogeneous
only on a scale $\sim D$, while to explain $W$ as the result of refractive
delay in a thin sheet would require much larger $\delta n_e$.  Hence we
consider only the bending of radiation, and not the difference between its
group velocity and $c$.
\section{Where Scattered?}
\label{where}
The inferred $\delta n_e$ (\ref{ne}) is more than seven orders of magnitude
greater than the maximum cosmologically allowable intergalactic $\langle n_e
\rangle \le 2 \times 10^{-7} (1 + z)^3\,\mathrm{cm}^{-3} = {\cal O}(10^{-6}
\,\mathrm{cm}^{-3})$.  A single scattering screen must have $\delta n_e
\gtrsim 10^7 \langle n_e \rangle$, a density much too great to be confined
in intergalactic space.

The pulse width might be explained as the result of ${\cal O} (10^{14})$
independent uncorrelated scatterings, each by a scatterer with $\delta n_e
\sim \langle n_e \rangle$.  The scattering regions must be $\lesssim
10^{14}$ cm in size.   There is no evident source of such fine scale
structure in the intergalactic medium, and it would be difficult to maintain
because at intergalactic densities the particle mean free paths are
${\cal O}(10^{18}T_{eV}^2\,$cm), much longer than the putative structure
size.  It would be smoothed rapidly by free particle flow, both of electrons
and of ions.  Henceforth we assume ${\cal O}$(1) scattering between emitter
and detection, rather than a large number of independent scatterings.

An additional argument against the hypothesis of intergalactic scattering is
that if scattering were distributed through the intergalactic medium, all
FRB should be broadened, with $W \propto D$ (\ref{delay}).  This is
inconsistent with the upper bounds on $W$ found \citep{T13} for three other
FRB.  The large inferred $\delta n_e$ requires that the pulse was scattered
in a dense localized region.

This cannot have been general intergalactic space.  The proportionality of
$W$ to $aD$, the distance of the scattering medium from the observer
(\ref{delay}), implies that scattering within the Galaxy produces roughly
the same pulse broadening of FRB as of Galactic pulsars (they differ by the
factor $(1-a)$, which is ${\cal O}(1/2)$ for typical Galactic sources, but
almost exactly unity for extragalactic sources).  The giant nanoshots of the
Crab pulsar exclude Galactic broadening of more than 0.4~ns at 9.25 GHz
\citep{HKWE03,HE07}, corresponding (with $\nu^{-4}$ scaling) to 800~ns
at 1400 MHz.  The nanoshots of PSR B1937+21 exclude Galactic broadening of
more than 15~ns at 1.65 GHz \citep{S04}, corresponding to 30~ns at 1400 MHz.
These upper bounds are negligible compared to the observed ms broadening of
some FRB, and exclude a significant Galactic contribution.

We consider scattering close to the source, writing $a = 1 - \epsilon$, with
$\epsilon \ll 1$.  Then (\ref{delay}) becomes
\begin{equation}
\label{theps}
\Delta \theta \approx \sqrt{2 c W \over \epsilon D} \approx (2
\times 10^{-10}, 3 \times 10^{-10}) \epsilon^{-1/2}.
\end{equation}
Assume single scattering and combine (\ref{neq}), (\ref{omegap}) and
(\ref{theps}):
\begin{equation}
\label{epsn2}
\epsilon \delta n_e^2 \approx {c W \omega^4 m_e^2 \over 2 \pi^2 D e^4}
\approx (70,150)\,\mathrm{cm}^{-6}.
\end{equation}
This also illustrates the familiar result $W \propto \omega^{\beta}$ with
$\beta = -4$ (consistent with pulsar data; \cite{B04}) independent of any
specific model of the distribution of $\delta n_e$, provided all the
structure occurs on scales $\gg \lambda/2\pi$ so that geometrical optics
applies.  Because $\Delta\theta \lesssim 1$ we can set a lower bound
(too small to be of interest) on the distance of the scatterer from the
source:
\begin{equation}
\label{epslower}
\epsilon D \gtrsim 2 c W \approx (3 \times 10^8,6 \times 10^8)\,\mathrm{cm}.
\end{equation}
\section{Dispersion Index}
\label{dindex}
The dispersion index $\alpha$, defined by the dispersion delay $\Delta t
\propto \nu^\alpha$, is a strong constraint on the density of the dispersing
plasma.  For FRB110220 $\alpha = -2.003 \pm 0.006$ \citep{T13} while for FRB
011025 $\alpha = -2.00 \pm 0.01$ \citep{BSB14}.  Expansion of the dispersion
relation for electromagnetic waves in a cold (nonrelativistic) plasma in
powers of $\omega_p^2/\omega^2 \ll 1$ yields
\citep{K14b}
\begin{equation}
\Delta t = \int\!{d\ell \over c}\,{1 \over 2}{\omega_p^2 \over \omega^2}
\left(1 + {3 \over 4}{\omega_p^2 \over \omega^2} + \cdots\right).
\end{equation}
Then
\begin{equation}
\label{alpha}
\alpha \equiv {d \ln{\Delta t} \over d \ln{\omega}} = -2 -{3 \over 2}
{\omega_p^2 \over \omega^2} + \cdots = -2 -{6 \pi n_e e^2 \over 
m_e \omega^2} + \cdots.
\end{equation}

In order to constrain $n_e$ in the scattering region we must allow for
the fact that it contributes only $\mathrm{DM_{scatt}}$ to the
(extra-Galactic) dispersion of the pulse.  The remainder, perhaps nearly
all, is attributed to intergalactic propagation, for which the higher terms
in (\ref{alpha}) are negligible.  From the observed bounds on $\alpha$,
(\ref{alpha}) yields
\begin{equation}
\label{DMlocal}
{\omega_p^2 \over \omega^2} {\mathrm{DM_{scatt} \over DM}} \le {2 \over 3}
\max{(-\alpha - 2)} = (0.006,0.007),
\end{equation}
where $\max{(-\alpha - 2)} \approx 0.01$ is the observed upper bound on
$-\alpha - 2$ for the FRB for which values are reported.

Using (\ref{omegap}), (\ref{epsn2}), (\ref{alpha}) and (\ref{DMlocal}),
\begin{equation}
\label{nebound}
n_e \approx \sqrt{(-\alpha - 2) c W \omega^6 m_e^3 \over 12 \pi^3 e^6
\mathrm{DM}} \lesssim (1.6 \times 10^8,2.6 \times 10^8)\ \mathrm{cm}^{-3},
\end{equation}
where the inequality results from the most negative values of $\alpha$
(-2.009, -2.01) permitted by the data.

The lower bound (\ref{ne}) on $\delta n_e$ may be combined with the upper
bound on $n_e$ implied by the maximum value of $(-\alpha -2)$ in
(\ref{nebound}), assuming $\delta n_e \sim n_e$, to yield a lower bound
\begin{equation}
\label{Dmin}
D \gtrsim {24 \pi e^2 \mathrm{DM} \over \max{(-\alpha -2)} \omega^2 m_e}
\sim 10^{14}\ \mathrm{cm};
\end{equation}
note that the scattering width $W$ drops out.  Ths bound is more than the
statement that the FRB occur outside the inner Solar System.

The electron density and the size $R$ of the dispersing region are bounded
from the plasma dispersion relation, without any consideration of the
scattering width:
\begin{equation}
R > {6 \pi \mathrm{DM} \over \max{(-\alpha -2)} \omega^2 m_e} \sim 2
\times 10^{13}\ \mathrm{cm}.
\end{equation}
This temperature-independent limit applies even to a perfectly homogeneous
plasma with no scattering at all, and excludes models that attribute the
dispersion to the immediate environment of a star.
\section{Fluctuation Density and Structure}
\label{parameters}
Make the plausible, but unproven, assumptions $\delta n_e \sim n_e$ and that
the same plasma disperses and scatters the FRB.
\subsection{Thin Screen}
A screen of index $n$ that refracts radiation may be very thin compared to
its distance from the source.  Its thickness (assuming transverse structure
on the same scale as its thickness) is limited by diffraction to
\begin{equation}
\label{deltah}
\Delta h \sim {\lambda \over 2 \pi \Delta \theta} \sim 3 \times 10^{10}
\epsilon^{1/2}\ \mathrm{cm}.
\end{equation}
Such a minimal thin screen contributes a negligible amount to DM:
\begin{equation}
\mathrm{DM}_{min} = \Delta h n_e \sim {c \nu m_e \over e^2} \sim 5 \times
10^{-8} \text{pc-cm}^{-3}.
\end{equation}
\subsection{Thick Plasma}
Assume a single scattering but that the electron density implied by the
scattering angle $\Delta \theta$ is present throughout the thickness
$\epsilon D$, even though the minimal $\Delta h$ (\ref{deltah}) may be
orders of magnitude smaller than $\epsilon D$.  Such a plasma could be an
outflowing wind from a point near the FRB source or from the FRB's
progenitor, provided (if from the progenitor) it is asymmetric so that the
propagation path is not parallel to its density gradient ($\tan\theta =
{\cal O}(1)$ in Eq.~\ref{Dthetan}).  Then
\begin{equation}
\label{DMlocalmax}
n_e \epsilon D \equiv \mathrm{DM_{local}} \le \mathrm{DM} = (910,680)\ 
\text{pc-cm}^{-3},
\end{equation}
where $\mathrm{DM_{local}}$ is the dispersion attributable to matter local
to the source that also causes the scattering.  Use
(\ref{epsn2}) to eliminate $n_e$, obtaining
\begin{equation}
\label{epsD}
\epsilon D \lessapprox {2 \pi^2 \mathrm{DM}^2 e^4 \over c W m_e^2 \omega^4}
\approx (1.3 \times 10^{13},4.4 \times 10^{12})\,\mathrm{cm};
\end{equation}
Combined with (\ref{Dmin}) this indicates that, whatever the distance to the
FRB, scattering occurs over a small fraction of that distance.
the corresponding density of the scattering matter
\begin{equation}
\label{nemin}
n_e \gtrsim \delta n_e \gtrapprox {c W m_e^2 \omega^4 \over 2 \pi^2
\mathrm{DM_{local}} e^4} \approx (2 \times 10^8,5 \times 10^8)\,
\mathrm{cm}^{-3}.
\end{equation}

The origin of these bounds on the dimensions and density of the plasma are
the large value of $W$, the assumptions of single scattering and of the
identity of the scattering and dispersing plasma.  The two bounds
(\ref{nebound}) and (\ref{nemin}) are slightly inconsistent for both FRB,
but because of the necessarily rough approximations made, this discrepancy
is not significant.  Their nearness does indicate that $n_e$ is near the
upper limit of the range allowed by (\ref{nemin}) and that a significant
fraction of the dispersion measure may be local to the source.

These limits correspond to $\mathrm{DM_{local} \approx DM}$, in which case 
$\mathrm{DM}$ cannot be used to infer the distance because an unknown
fraction, perhaps nearly all, of the dispersion is local to the source.
For FRB for which only upper bounds on $W$ exist, there is neither a lower
bound on $n_e$ nor an upper bound on $\epsilon D$.  If the scattering plasma
contributes only a fraction of the dispersion, then DM should be taken only
as that fraction of the total, further tightening the bounds.

If we take the lower bound (\ref{epslower}) on $\epsilon D$ rather than
the upper bound (\ref{epsD}) then, using (\ref{epsn2}), 
\begin{equation}
\label{dnemax}
\delta n_e \lessapprox \omega^2 m_e/(2 \pi e^2) \approx 4 \times 10^{10}\
\mathrm{cm}^{-3},
\end{equation}
independent of the parameters of any particular FRB.  This amounts, aside
from a factor of two, to the condition that the radiation propagate through
the scattering plasma.  If $n_e$ approaches this bound then $\omega_p
\approx \omega$ and $\Delta \theta = {\cal O}(1)$.  Such a dense cloud may
also have been the source of the FRB emission \citep{K14a}, but cannot be a
major contributor to the dispersion measure because of the arguments made
in \S~\ref{dindex}.

The limits (\ref{epslower}) and (\ref{nemin}) imply
\begin{equation}
\label{DMlocalmin}
\mathrm{DM_{local}} \gtrapprox {c^2 W^2 \omega^4 m_e^2 \over \pi^2
\mathrm{DM} e^4} \approx (0.02,0.09)\, \text{pc-cm}^{-3},
\end{equation}
consistent with the lower bound (\ref{DMlocal}) on $\mathrm{DM_{scatt}}$.
The two limits (\ref{DMlocalmax}) and (\ref{DMlocalmin}) bound the possible
range of $\mathrm{DM_{local}}$, and correspond to the bounds (\ref{nemin})
and (\ref{dnemax}) on $\delta n_e$.  The local contribution to the
dispersion measure may, but need not, be very small.
\section{Constraints on FRB Models}
\label{implications}
The results of this paper impose a number of constraints on the astronomical
environments in which FRB are produced:
\subsection{Number of FRB Sources} \label{Nsources}
There are two constraints on the number of presently active detectable FRB
sources $N_{sources} \equiv BT$, where $B$ is their birth rate within the
volume from which FRB may be detected and $T$ is their active lifetime
(consistent with the known properties of FRB, such as their dispersion
measures).  If the bursts occur stochastically, without any latency period
following a burst, then the absence of coincidences among $N_{FRB}$
observed FRB implies
\begin{equation}
\label{norep}
N_{sources} \gtrsim N_{FRB}^2.
\end{equation}

The absence of repetitions of any individual FRB implies
\begin{equation}
\label{nsources}
N_{sources} \gtrsim \Omega_{FRB} \tau_{min} \sim 10^4,
\end{equation}
where $\Omega_{FRB}$ is the all-sky FRB rate and $\tau_{min}$ is the
empirical lower bound on the repetition time of an individual source.
\cite{T13} estimate $\Omega_{FRB} \sim 0.3$/s while \cite{Ku14} estimate
$\Omega_{FRB} \sim 0.1$/s; the spread between these two values is an
indication of their uncertainty.

If the bursts are stochastic then $\tau_{min} \sim \tau_{tot}$, the total
time beams pointed in the {\it known\/} directions to FRB, summed over all
FRB, without observing a repetition\footnote{It is not necessary that a beam
be pointed to a single FRB for this time because, if they all have the same
properties, staring in any direction in which an FRB has been observed is
equivalent.  It is also assumed that localization is good enough that the
chance of misidentifying a new source as a repetition of a previously
observed source is negligible; for $15^\prime$ localization and $N_{FRB}
\sim 10$ this chance is $\sim 10^{-5}$.}.  \cite{L14} found no recurrences
in $1.1 \times 10^5$ s of observations of a single FRB, implying a 95\%
confidence bound $\tau_{min} > 2.7 \times 10^4$ s, giving the numerical
estimate in (\ref{nsources}).  On the other hand, if there is a latency
period between FRB from a single source then, depending on how observing
time was distributed, $\tau_{min}$ may be as short as $\tau_{cont}$, the
longest duration of continuous observation of an individual FRB location
without a repetition.

The conditions (\ref{norep}) and (\ref{nsources}) may be used to test models
of $N_{sources}$ against the empirical parameters $N_{FRB}$, $\tau_{min}$
and $\Omega_{FRB}$, and thereby to constrain models of the sources, of their
astronomical environments, and of their distances.  If more than one FRB
were observed from the same direction then (\ref{norep}), with the right
hand side divided by the number of coincidences, would become an approximate
equality.
\subsection{Supernovae, Soft Gamma Repeaters and Their Remnants}
\label{SN}
The discovery \citep{K12,S14b} of two apparent FRB at low Galactic latitude
suggests they may be cosmologically local and associated with our Galaxy.
\cite{Ku14} suggest an association with the giant flares of
SGR, with dispersion originating in the surrounding young SNR, and a lower
bound on FRB distances of 300 kpc (for an assumed temperature of the
dispersive plasma of $8000^{\,\circ}$K).  The local dispersion measure of a
source at the center of a spherical cloud of ionized gas of mass $M$ and
radius $R$ is
\begin{equation}
\label{SNRDM}
\mathrm{DM_{local}} = 818 {M \over M_\odot} \left({R \over 0.1\,\mathrm{pc}}
\right)^{-2} f\ \mathrm{pc\mbox{-}cm^{-3}},
\end{equation}
where $f = 1$ for a homogeneous sphere and $f = 1/3$ for a thin shell,
implying $R \sim 0.1\,$pc for a SNR, lost stellar envelope, {\it etc.\/},
that provides much of the dispersion measure of a FRB.  Note, however, that
by (\ref{epsD}) scattering by such a cloud cannot also explain the observed
pulse widths.

The age and lifetime $T$ of an expanding cloud
\begin{equation}
\label{SNRlife}
T \approx {R \over V} \approx 30 {R \over 0.1\,\mathrm{pc}} {3000\,
\mathrm{km/s} \over V}\ \mathrm{y} \approx 30 \sqrt{{f \over
\mathrm{DM}_{1000}}{M \over M_\odot}} {3000\,\mathrm{km/s} \over V}\ 
\mathrm{y},
\end{equation}
where $V$ is the expansion velocity and $\mathrm{DM}_{1000} \equiv
\mathrm{DM}/(1000\,\text{pc-cm}^{-3})$.  At $R = 0.1\,$ pc only $\sim
10^{-4} n_{ISM} M_\odot$ of interstellar material will have been swept up,
for an interstellar density of $n_{ISM}$ atoms/cm$^3$, so $V$ is nearly the
initial explosion velocity.  If $V$ is within the range 3000--30000 km/s of
SN ejecta then the age of the dispersing cloud $T \lesssim 30$ y.  If FRB
are found within such clouds, then if repetitive bursts are observed their
dispersion measures will decrease monotonically and smoothly according to
(\ref{SNRDM}) with $R = Vt$.  The hypothesis that the dispersion is produced
by very young cosmologically local SNR is contradicted by the absence of SN
within the last $\sim 30$ y at the high Galactic latitudes of most FRB.

The number of SNR with ages $t < T$ (\ref{SNRlife}) associated with our
Galaxy (out to distances $\sim 1$~Mpc) is inferred from the SN rate to be
$N_{SNR\ t < T} \lesssim {\cal O}$(1).  The hypothesis that the dispersion
measures of FRB result from propagation through such young and nearby SNR is
also contradicted by the fact that no repeaters are observed among seven FRB
when only $\lesssim {\cal O}(1)$ SNR young enough to meet this requirement
likely exists within 1 Mpc.  Further, the all-sky FRB rate $\Omega_{FRB}
\sim 0.1$--0.3/s  would imply a repetition time of an individual source
$\tau \sim N_{sources}/\Omega_{FRB} = N_{SNR\ t < T}/ \Omega_{FRB} \sim
3$--10\,s.  The hypothesis of such rapid repetitions of FRB is excluded
empirically \citep{L14}.

If FRB are associated with SN, at a rate of order one-to-one (the FRB do not
repeat), comparison of the rates of the two classes of events shows that
their distances must be cosmological:  The SN rate is estimated \citep{S07}
to be $\Omega_{SN} \approx 0.098 \times 10^{-12} M_\odot^{-1}$-y$^{-1}$.
Standard cosmological parameters indicate a local baryon density
$\rho_{baryon} = 1.9 \times 10^{-64} M_\odot$ cm$^{-3}$ and a SN rate
$\Omega_{SN} \rho_{baryon} \approx 1.9 \times 10^{-77}$ cm$^{-3}$-y$^{-1}$.
Comparison to the all-sky FRB rate $\Omega_{FRB} \approx 0.1$--0.3\,/s
indicates that SN out to a distance of $\sim 1\,$Gpc must contribute.
Unless the volumetric FRB rate is much higher than
the SN rate, as might be the case if FRB are giant pulsar pulses (excluded
by their dispersion measures, unless at cosmological distances), SGR
outbursts \citep{Ku14}, or other phenomena that repeat many times in their
sources' lifetimes, FRB originate at cosmological distances, even if much of
their dispersion measures is local to their sources.

If, on the other hand, many FRB are associated with each SN, we can set a
lower bound on the distance out to which FRB are observed:
\begin{equation}
D \gtrsim \left({3 N_{sources} \over 4 \pi \Omega_{SN} \rho_{baryon}T_{FRB}}
\right)^{1/3} \sim 10\ \mathrm{Mpc},
\end{equation}
where $T_{FRB}$ is the FRB-active lifetime of the remnant of a SN; the
numerical value assumes $T_{FRB} \sim 3000$ y, the estimated active lifetime
of a SGR.  The absence of obvious correlation with cosmologically local
structure such as the Coma cluster suggests $D \gtrsim {\cal O}(100)$ Mpc.
\subsection{Inverse Bremsstrahlung}
\label{invbrems}
If $\delta n_e \sim n_e$ and the density $n_e$ is found over a path length
$\Delta h$ then $\Delta h = \epsilon D$ and (\ref{epsn2}) imply an inverse
bremsstrahlung optical depth $\tau_{ff} \propto \epsilon D n_e^2$ in the
scattering medium, independent of the particular values of $\epsilon$ and
$\delta n_e$.  Aside from the medium temperature, this depends only on
observed quantities:
\begin{equation}
\label{tauff}
\tau_{ff} \approx {4 \over 3} \sqrt{2 \pi \over 3 k_B T} {n_e^2 \epsilon D
e^6 \over k_B T c m_e^{3/2} \nu^2} g_{ff} = {8 \over 3} \sqrt{2 \pi m_e
\over 3 k_B T} {W \omega^2 e^2 \over k_B T} g_{ff} \approx (2.3,3.9) \left(
{10^{7\,\circ} \mathrm{K} \over T}\right)^{3/2},
\end{equation}
where the Gaunt factor $g_{ff} \approx 11.5$ \citep{S62}.  In order that
$\tau_{ff} \lesssim 1$ it is necessary that either $T \gtrsim
10^{7\,\circ}$K or $\langle \delta n_e^2 \rangle \gg \langle n_e\rangle^2$
(the scattering matter be a thin dense screen) in a much more dilute medium.
The first possibility is consistent with a region of high energy density;
the second is also possible but would vitiate the assumption $\delta n_e
\sim n_e$.  The condition (\ref{deltah}) is consistent with very thin
screens, as in some models (Section \ref{PSRJ1745}) must be responsible for
the scattering of PSR J1745-2900.  Even if $W$ is not measured, $\tau_{ff}
\lesssim 1$ still imposes a temperature-dependent constraint on the emission
measure $\int n_e^2 \, d\ell = \int n_e^2 D \,d\epsilon$ along the path
between the source and the observer \citep{Ku14}.
\subsection{Jeans Limit}
\label{Jeans}
If the dispersion occurs in a stable {\it static\/} plasma cloud, then the
Jeans condition that the cloud be stable against gravitational collapse
imposes further constraints on its parameters:
\begin{equation}
\sqrt{GM \over R} \lessapprox c_s = \sqrt{5 k_B T (1 + \mu) \over 3 m_p},
\end{equation}
where $c_s$ is the sound speed and $\mu \approx 0.85$ is the number of 
electrons per baryon.  Substituting $M \approx R^3 m_p n_e/\mu$ and
$\mathrm{DM} \approx n_e R$ (attributing the dispersion to the source's
plasma cloud, not the the intervening line of sight), we find
\begin{equation}
\label{RmaxJ}
R \lessapprox {5 (1 + \mu) \mu k_B T \over 3 G \mathrm{DM} m_p^2} \approx
5 \times 10^{21} {T_{8000} \over \mathrm{DM}_{1000}}\ \mathrm{cm}
\end{equation}
and
\begin{equation}
\label{neminJ}
n_e \sim {\mathrm{DM} \over R} \gtrapprox 0.6\,\mathrm{DM}_{1000}^2
T_{8000}^{-1}\ \mathrm{cm}^{-3},
\end{equation}
where we normalize the temperature $T_{8000} \equiv T/8000^{\,\circ}$K
(following \cite{Ku14}) and the dispersion measure $\mathrm{DM}_{1000}
\equiv \mathrm{DM}/$1000 pc-cm$^{-3}$, and assume complete ionization and
cosmic abundances.  The corresponding mass 
\begin{equation}
\label{MmaxJ}
M \lessapprox {25 k_B T (1 + \mu)^2 \mu \over 9 G^2 m_p^3 \mathrm{DM}}
\approx 8 \times 10^7 {T_{8000}^2 \over \mathrm{DM}_{1000}}\ M_\odot.
\end{equation}
The hydrodynamic time
\begin{equation}
T_J \sim {R \over c_s} \lessapprox \sqrt{5 k_B T (1+\mu) \over 3 m_p} {\mu
\over G m_p \mathrm{DM}} \approx 10^8 {T_{8000}^{1/2} \over
\mathrm{DM}_{1000}}\ \mathrm{y}
\end{equation}
has no explicit dependence on the unknown parameters $n_e$, $R$ and $M$.
$T_J$ is long enough to avoid the statistical problems (Section \ref{SN})
posed by attributing the dispersion measures to young Galactic SNR, whose
youth implies that only a very few are active with the observed dispersion
measures at any time.  The dispersive cloud could be more compact and dense
than the bounds (\ref{RmaxJ}) and (\ref{neminJ}), perhaps by a large factor.

These bounds are consistent with dense static compact clouds in the Galactic
neighborhood while avoiding the rapid expansion and short lifetime implied
by attributing them to rapidly expanding young SNR.  Much smaller $R$ and
$M$ and larger $n_e$ than the bounds are possible.  The bounds also admit a
protogalaxy or starburst ionized by an initial generation of hot luminous
stars, providing the observed dispersion measures.  Such sites may be
plausible locales for FRB, but give no clues to the origin of the FRB
themselves beyond indicating a relation with massive stars and high rates of
star formation and death.  As argued in Sections \ref{where} and
\ref{parameters}, these clouds cannot be the origin of the observed pulse
widths, but may contribute a major part of the total dispersion measures.
\section{Discussion}
\label{discuss}
The central results (\ref{ne}), (\ref{epsn2}) and (\ref{epsD}) of this
paper are that the pulses of FRB 110220 and 011025, the two FRB with pulse
widths attributed to scattering, scattered in high density regions close to
their sources.  We also infer, from the closeness of the dispersion indices
to their low density value of $-2$, that dispersion occurred in a region
where the electron density was close to the bound (\ref{nemin}) and that a
significant part of the dispersion occurred close to the source.  The
distances inferred from the dispersion measures are then only upper bounds,
although the fact that most FRB occurred at high Galactic latitudes implies
that they are either extra-Galactic or very close ($\lesssim 100\,$pc).

These results depend on the assumption $\delta n_e \sim n_e$.  This
assumption could be violated in many ways.  For example, the pulse width
might have been produced by the reverberation of radio emission in a cavity
of size $< cW$ if the walls of the cavity had a plasma density above the
critical density $n_e \approx 2.1 \times 10^{10}\,$cm$^{-3}$ for 1300 MHz
radiation and the interior had a lower, perhaps much lower, density.
However, reverberation would be unlikely {\it a priori\/} to produce a
scattering index $\beta \approx -4$.  Alternatively, if scattering occurs in
a thin comparatively dense sheet $\delta n_e \gg \langle n_e \rangle_{LOS}$,
where $\langle n_e \rangle_{LOS} \equiv \mathrm{DM}/D$, and there may be
evidence for such sheets in our Galaxy (\ref{PSRJ1745}, \cite{B14}).

This paper began by assuming that the FRB are at the cosmological distances
inferred from their dispersion measures, allowing only for the estimated
Galactic dispersions \citep{T13}.  As shown in Section~\ref{dindex}, this is
only marginally consistent with the dispersion indices.  The fact that for
both FRB (110220 and 011025) whose pulse widths are attributed to scattering
the consistency between (\ref{nebound}) and (\ref{nemin}) is only marginal
should be of concern.  It is {\it a priori\/} surprising that both objects
should be found in the same corner of the allowable parameter space, the
range of plasma densities allowed by the pulse widths, which hints
at a fundamental problem with the model.

This suggests that for some, as yet undiscovered, FRB, either a significant
deviation from the low density plasma dispersion index $\alpha = -2$ will be
found, or there will be a frank inconsistency between the observed $\alpha$
and that inferred from (\ref{alpha}) and (\ref{nebound}).  Such an
inconsistency may require reconsideration of the interpretation of the
pulse widths as the effects of scattering or of the dispersion measures as
indicating cosmological distances, as \cite{BSB14} and \cite{Ka14} have done
on other grounds.  If so, the distances are smaller than inferred from the
dispersion measures, perhaps by large factors.

If we reject the inference of cosmological distances then various bounds
change.  The lower bound (\ref{ne}) on $\delta n_e$ scales $\propto
D^{-1/2}$, the estimate (\ref{epsn2}) of $\epsilon \delta n_e^2$ scales
$\propto D^{-1}$, but the bounds on density (\ref{DMlocal}), (\ref{nebound})
and (\ref{nemin}) are independent of $D$.  At $D \sim 30\,$kpc (\ref{ne})
becomes $\delta n_e \gtrsim 6 \times 10^3\,$cm$^{-3}$, consistent with a
young SNR \citep{Ku14}.  The bounds (\ref{nebound}), (\ref{Dmin}) and
(\ref{tauff}) exclude origin in local plasmas, such as meteor trails (the
dates of FRB reported by \cite{T13} do not coincide with meteor showers),
lightning and electric discharges.

Finally, we note that radar systems use chirped emission, compressed upon
reception into narrow pulses, in order to obtain accurate range measurements
without requiring excessive peak transmitted powers.  The observation of FRB
in a single beam at Parkes, in contrast to perytons \citep{BS11}, indicates
a distance $\gtrsim 20$ km, outside the first Fresnel zone, consistent with
a radar satellite.  There is no obvious reason for a radar to have a chirp
$\omega \propto t^{-1/2}$ as observed, nor is there obvious reason not.
However, the observed dispersed pulse durations of several tenths of a
second would imply, for monostatic radar, target distances of at least half
that many light seconds to avoid interference of the transmission with the
received scattered radiation.  At such distances $\sim 10^{10}\,$cm the
return would be undetectably weak.  In contrast, bistatic radar can use
arbitrarily long pulses.  The pulse repetition intervals would have to have
been longer than the lengths of time the radars were anywhere in the 13
beams of the Parkes Multibeam Pulsar Survey (about 0.3 s for a radar in low
Earth orbit moving perpendicularly to a beam), yet the pulse durations must
have been shorter than the time required to cross a single beam.  This
explanation would also require at least as many radar satellites, each with
a different chirp rate, as FRB because each FRB had a different dispersion
measure, or satellites whose chirp rates were variable in some non-obvious
manner.  This combination of requirements makes the hypothesis of
interference by an orbital chirped source implausible.

I thank T. Piran for useful discussions.
\appendix
\section{Application to Galactic PSR}
\label{others}
In this appendix we apply the preceding results to some Galactic pulsars
for which $W$ is constrained empirically.
\subsection{Crab PSR and PSR B1937+21}
Both these objects show nanoshots from which upper limits can be placed on
broadening by scattering in the plasma on the line of sight.  The Crab PSR
has nanoshots of width $\le 0.4$ ns at 9.25 GHz \citep{HE07}, implying $W
\lesssim 0.4$ ns.  This sets an upper limit to the lower bound (\ref{ne}):
$n_e \gtrsim n_{e,min}$ where $n_{e,min} \lesssim 260$ cm$^{-3}$.  Because
the actual $n_{e,min}$ cannot be estimated (only a bound on $W$ exists, not
a measured value), it is not possible to infer anything about $n_e$ from the
scattering argument.  The dispersion measure does constrain $n_e$, but is
consistent with, for example, a perfectly homogeneous medium with no
scattering at all.  A similar argument for PSR B1937+21, with broadening
$\le 15$ ns at 1.65 GHz \citep{S04} leads to $n_{e,min} \lesssim 40$
cm$^{-3}$.

The actual values of $n_{e,min}$ may be orders of magnitude less, making
them consistent with interstellar plasma densities.  This argument can be
inverted to predict $W$ from known properties of the interstellar medium,
with the conclusion that $W$ is several orders of magnitude less than the
present empirical upper limits.
\subsection{PSR J1745-2900}
\label{PSRJ1745}
PSR J1745-2900 at the Galactic Center has $W = 1.3$ s at 1 GHz (fitted to  
observations at a range of frequencies from 1.2 GHz to 18.95 GHz with a
power law with scattering index consistent with -4) and $\mathrm{DM} =
1778$ pc-cm$^{-3}$ \citep{S14a}.  Its line of sight passes within about
$3^{\prime\prime}$ (0.1 pc) of Sgr A$^\mathrm{*}$.  This is statistically
unlikely to be coincidental, and suggests a physical association within that
distance of the massive black hole.

We consider the hypothesis that much of the extraordinary scattering and
dispersion measure of PSR J1745-2900 are associated with the immediate
environment of Sgr A$^\mathrm{*}$, so that $\epsilon D = 0.1$ pc ($\epsilon
\approx 1.2 \times 10^{-5}$).  This hypothesis is the natural explanation of
the fact that its pulse broadening is several orders of magnitude greater
than those of other PSR, such as the Crab PSR and PSR B1937+21.  If this
broadening were the result of scattering in the general interstellar medium,
it would be expected to be roughly comparable for all PSR at comparable
distances, not differ by orders of magnitude; the location of PSR J1745-2900
at the Galactic Center is extraordinary, but its propagation path through
the interstellar medium is not.

Taking $\epsilon D = 0.1$ pc and using (\ref{epsn2}), we find $\delta n_e
\approx 1.2 \times 10^7$ cm$^{-3}$.  However, $\epsilon D \delta n_e \approx
1.2 \times 10^6$ pc-cm$^{-3}$, nearly 1000 times greater than the actual
dispersion measure (some of which must be attributed to the 8.3 kpc path
through the interstellar medium).  From this we infer that the scattering
occurs in a thin screen whose thickness $\Delta h \lesssim 4 \times 10^{14}$
cm (a fractional thickness $\Delta h/\epsilon D \lesssim 10^{-3}$).  Using 
(\ref{alpha}) we predict a dispersion index
\begin{equation}
\alpha = -2 - 0.0015 \left({1\,\mathrm{GHz} \over \nu}\right)^2 + \cdots.
\end{equation}
The coefficient of the $\nu^{-2}$ term is uncertain because the geometry is
uncertain.  Its measurement would be the first demonstration in an
astronomical context of the higher terms in (\ref{alpha}).

However, the assumption $\epsilon D \sim 0.1$ pc implies an angular size,
(Fig.~\ref{pathfig}) using (\ref{theps}), $\phi \approx \epsilon \Delta
\theta \approx 1 (\nu/\text{1 GHz})^{-2}$ mas, in contradiction to the
measured angular size, extrapolated to 1 GHz, of 900 mas \citep{B14}.  We
therefore reject the hypothesis of thin screen scattering with $\epsilon D
= 0.1$ pc. 

By comparing the pulse broadening and angular size \cite{B14} concluded that
the scattering screen is actually $5.8 \pm 0.3$ kpc from the Galactic Center
($a = 0.3$); the special environment surrounding Sgr A$^\mathrm{*}$
contributes little.  Using (\ref{ne}), we find $\delta n_e \approx 9 \times
10^4$ cm$^{-3}$, an extraordinary electron density for interstellar space.
If, rather than a single scattering sheet there are $N$ such sheets,
$\delta n_e$ is reduced by a factor ${\cal O}(N^{-1/2})$.

The requirement (\ref{tauff}) $\tau_{ff} \le 1$ implies a sheet thickness
$\Delta h \lessapprox 7 \times 10^{14} T_{eV}^{3/2} (\nu/\text{1 GHz})^2$ cm
(because $\delta n_e \sim n_e \propto N^{-1/2}$ this result is independent
of $N$) and a contribution to the dispersion measure $\le 20$ pc-cm$^{-3}$.
The thinness of this sheet, and the problem that it is found on the line of
sight to PSR J1745-2900 at the Galactic Center but nothing like it is found
on lines of sight to other pulsars, is disturbing.  The characteristic
expansion time $\Delta h/v_{th} \le 20 T_{eV}$ y, suggesting that changes in
the pulse broadening, angular size, and possibly DM (more accurately
measurable, though proportionately smaller) may be observable over a decade,
offering the possibility of an independent test.

Each of the hypotheses for the location of a thin scattering screen meets
serious objections:  The hypothesis that it is within $\sim 0.1$ pc of the
Galactic Center and of the pulsar implies an angular size nearly three
orders of magnitude smaller than observed.  The hypothesis that it is in the
general interstellar medium at the distance (6 kpc from the Galactic Center
and 2 kpc from us) that reconciles the pulse broadening and angular width
demands an uncomfortably high screen density and fails to explain why this
line of sight, far from the unique Galactic Center, is special: lines of
sight to other pulsars don't intersect screens with similar parameters, as
shown by the observation that their pulses are broadened (at 1 GHz) by
$< 1\ \mu$s, compared to about 1 s for PSR J1745-2900.  

In order to resolve this problem (the pulse broadening of PSR J1745-2900 is
too short for its angular size; equivalently, its angular size is too wide
for its pulse broadening) we suggest that the radiation is refracted by
plasma near its source, and that we are near a caustic or a focal point.
This requires rejection of the {\it ad hoc\/} thin screen scattering model,
in which there is no correlation between the scattering directions of
adjacent points on the screen, and scattered radiation is delayed compared
to unscattered radiation, or to that scattered by smaller angles.  In
contrast, all rays converging on a caustic or optical focus have the same
travel time from the source because, by Fermat's Principle that travel time
is an extremum, convergence is possible only for rays that have the same
(extremum) travel time.  This argument assumes the travel time is a
continuous function of angular displacement (the lens properties must be
continuous functions of displacements perpendicular to the rays), and does
not apply to Fresnel lenses.

The rays may have an arbitrarily broad angular distribution that depends
on the geometry of the refracting medium.  Sources for which we are on a
caustic or at a focus are brightened, perhaps explaining why the sole
detected Galactic Center pulsar satisfies this unusual and unexpected
condition.  The observation of nonzero scattering widths for
PSR J1745-2900 with scattering index $\beta = -3.8 \pm 0.2$ \citep{S14a}
indicates that there is a contribution to the detected radiation that is
described by the thin sheet scattering model.

These arguments cannot be applied to FRB because no measurements of their
angular sizes exist.  Interferometric detection of FRB \citep{L14} might
permit measurements of their angular sizes, and be a critical test.  A more
quantitative investigation is the subject of ongoing work, but is beyond the
scope of this paper.


\begin{thebibliography}{99}
\bibitem[\protect\citeauthoryear{Bhat, {\it et al.\/}}{2004}]{B04} Bhat,
N. D. R., Cordes, J. M., Camilo, F., Nice, D. J. \& Lorimer, D. R. 2004
\apj\ 605, 759.
\bibitem[\protect\citeauthoryear{Bower, {\it et al.\/}}{2014}]{B14} Bower,
G. C., {\it et al.\/} 2014 \apjl\ 780, L2.
\bibitem[\protect\citeauthoryear{Burke-Spolaor, {\it et al.\/}}{2011}]{BS11}
Burke-Spolaor, S., Bailes, M., Ekers, R., Marquart, J.-P. \& Crawford,
F.~III 2011 \apj\ 727, 18.
\bibitem[\protect\citeauthoryear{Burke-Spolaor \& Bannister}{2014}]{BSB14}
Burke-Spolaor, S. \& Bannister, K.~W. 2014 arXiv:1407.0400.
\bibitem[\protect\citeauthoryear{Hankins \& Eilek}{2007}]{HE07}
Hankins, T.~H. \& Eilek, J.~A. 2007 \apj\ 670, 693.
\bibitem[\protect\citeauthoryear{Hankins, {\it et al.\/}}{2003}]{HKWE03}
Hankins, T.~H., Kern, J.~S., Weatherall, J.~C. \& Eilek, J.~A. 2003 \nat\
422, 141.
\bibitem[\protect\citeauthoryear{Karbelkar}{2014}]{Ka14} Karbelkar, S.~N.
2014 arXiv:1407.5653.
\bibitem[\protect\citeauthoryear{Katz}{2014a}]{K14a} Katz, J. I. 2014a \prd\
89, 103009.
\bibitem[\protect\citeauthoryear{Katz}{2014b}]{K14b} Katz, J. I. 2014b \apj\
788, 34.
\bibitem[\protect\citeauthoryear{Keane, {\it et al.\/}}{2012}]{K12} Keane,
E.~F., Stappers, B.~W., Kramer, M. \& Lyne, A.~G. 2012 \mnras\ 425, L71.
\bibitem[\protect\citeauthoryear{Kulkarni, {\it et al.\/}}{2014}]{Ku14}
Kulkarni, S.~R., Ofek, E.~O., Neill, J.~D., Zhang, A. \& Juric, M. 2014
\apj\ 797, 70.
\bibitem[\protect\citeauthoryear{Law, {\it et al.\/}}{2014}]{L14} Law, C.
J., {\it et al.\/} 2014 arXiv:1412.7536.
\bibitem[\protect\citeauthoryear{Lorimer, {\it et al.\/}}{2007}]{L07}
Lorimer, D. R., Bailes, M., McLaughlin, M. A., Narkevic, D. J., \& Crawford,
F. 2007 Science 318, 777.
\bibitem[\protect\citeauthoryear{Sharon, {\it et al.\/}}{2007}]{S07}
Sharon, K., Gal-Yam, A., Maoz, D., Filippenko, A.~V. \& Guhathakurta, P.,
2007 \apj\ 660, 1165.
\bibitem[\protect\citeauthoryear{Soglasnov, {\it et al.\/}}{2004}]{S04}
Soglasnov, V.~A., Popov, M.~V., Bartel, N., Cannon, W., Novikov, A.~Y.,
Kondratiev, V.~I. \& Altunin, V.~I. 2004 \apj\ 616, 439.
\bibitem[\protect\citeauthoryear{Spitler, {\it et al.\/}}{2014a}]{S14a}
Spitler, L. G. {\it et al.\/} 2014a \apjl\ 780, L3.
\bibitem[\protect\citeauthoryear{Spitler, {\it et al.\/}}{2014b}]{S14b}
Spitler, L.~G. {\it et al.\/} 2014b \apj\ 790, 101.
\bibitem[\protect\citeauthoryear{Spitzer}{1962}]{S62} Spitzer, L. {\it
Physics of Fully Ionized Gases\/} 2nd ed. (Interscience, New York 1962).
\bibitem[\protect\citeauthoryear{Thornton, {\it et al.}}{2013}]{T13}
Thornton, D., {\it et al.\/} 2013 Science 341, 53.
\bibitem[\protect\citeauthoryear{Williamson}{1972}]{W72} Williamson, I. P.
1972 \mnras\ 157, 55.
\end{thebibliography}
\end{document}